\documentclass[aps, floatfix, nofootinbib, amsmath]{revtex4}
\usepackage{times}

\usepackage{braket}
\usepackage{enumerate}
\usepackage{subfigure}
\usepackage{bm}
\usepackage{epsfig}
\usepackage{epstopdf}
\usepackage{subfigure}
\usepackage{graphicx} 
\usepackage{amsmath}
\usepackage{amssymb}
\usepackage{soul}
\usepackage[T1]{fontenc}
\usepackage{hyperref}
\hypersetup{
 colorlinks = true,
 linkcolor = blue,
 citecolor = blue,
 urlcolor = blue
}

\usepackage{hyperref}
\usepackage{url}

\newcommand{\BEA}{\begin{eqnarray}}
\newcommand{\EEA}{\end{eqnarray}}
\renewcommand{\d}{{\rm d}}
\newcommand{\D}{{\cal D}}
\newcommand{\comment}[1]{}

\begin{document} 

\title{Maximum Entropy competes with Maximum Likelihood}

\author{A.E. Allahverdyan \& N.H. Martirosyan}

\affiliation{A. Alikhanyan National Laboratory (Yerevan Physics Institute), 0036 Yerevan, Armenia}

\begin{abstract}
Maximum entropy (MAXENT) method has a large number of
applications in theoretical and applied machine learning, since it
provides a convenient non-parametric tool for estimating unknown
probabilities.  The method is a major contribution of statistical
physics to probabilistic inference. However, a systematic approach
towards its validity limits is currently missing.  Here we study MAXENT
in a Bayesian decision theory set-up, i.e. assuming that there exists a
well-defined prior Dirichlet density for unknown probabilities, and that
the average Kullback-Leibler (KL) distance can be employed for deciding
on the quality and applicability of various estimators.  These allow to
evaluate the relevance of various MAXENT constraints, check its general
applicability, and compare MAXENT with estimators having various degrees
of dependence on the prior, {\it viz.} the regularized maximum
likelihood (ML) and the Bayesian estimators. We show that MAXENT applies
in sparse data regimes, but needs specific types of prior information. 
In particular, MAXENT can outperform the optimally regularized
ML provided that there are prior rank correlations between the estimated
random quantity and its probabilities. 

\end{abstract}

\maketitle

\newcommand{\fix}{\marginpar{FIX}}
\newcommand{\new}{\marginpar{NEW}}


\section{Introduction}

The maximum entropy (MAXENT) method was proposed within statistical
physics \citep{jaynes,balian,rmp}, and later on got a wide
range of inter-disciplinary applications in data science, probabilistic
inference, biological data modeling {\it etc}; see e.g.  \citep{volume}.
MAXENT estimates unknown probabilities (that generated data) via
maximizing the Boltzmann-Gibbs-Shannon entropy under certain constraints
which can be derived from the observed data \citep{volume}. MAXENT leads
to {\it non-parametric} estimators whose form does not depend on the
underlying mechanism that generated data (i.e.  prior assumptions).
Also, MAXENT avoids the zero-probability problem, i.e.  when operating
on a sparse data, so that certain values of the involved random quantity
may not appear due to a small, but non-zero probability, MAXENT still
provides a controllable non-zero estimate for this small probability. 

MAXENT has has several {\it formal} justification
\citep{jaynes,chakra,baez,cover,topsoe,shore,paris}. But the following
open problems are basic for MAXENT, because their insufficient
understanding prevents its {\it valid} applications. {\it (i)} Which
constraints of entropy maximization are to be extracted from data, which
is necessarily finite and noisy? {\it (ii)} When and how these
constraints can lead to overfitting, where, due to a {\it noisy}
data, involving more constraints leads to poorer results?  {\it (iii)}
How predictions of MAXENT compare with those of other estimators, e.g.
the (regularized) maximum likelihood? 

Here we approach these open problems via tools of Bayesian decision
theory \citep{cox}.  We assume that the data is given as an i.i.d.
sample of a {\it finite} length $M$ from a random quantity with $n$
outcomes and unknown probabilities that are instanced from a non-informative prior Dirichlet
density, or a mixture of such densities.  Focusing on the sparse data
regime $M<n$ we calculate average KL-distances between real probabilities
and their estimates, decide on the quality of MAXENT under various
constraints, and compare it with the (regularized) maximum-likelihood
(ML) estimator.  Our main results are that MAXENT does apply to
sparse data, but does demand specific prior information. We
explored two different scenarios of such information. First, the unknown
probabilities are most probably deterministic. Second, there are
prior rank correlations between the inferred random quantity and its
probabilities. Moreover, in the latter case the non-parametric MAXENT
estimator is better in terms of the average KL-distance than the {\it
optimally} regularized ML (parametric) estimator.

Some of above questions were already studied in literature.
\citep{good_entropy,christ,minent,mdl} applied formal principles of statistics
(e.g. the Minimum Description Length) to the selection of constraints
(question {\it (i)}). Our approach to studying this question will be
direct and unambiguous, since, as shown below, the Bayesian decision theory leads
to clear criteria for the validity of MAXENT estimators. We can
also compare all predictions with the optimal Bayesian estimator. The
latter is normally not available in practice due to insufficient
knowledge of prior details, but it still does provide an important
theoretical benchmark.  Note that
\citep{thomas,lebanon,kazama,smola,dudik,rau,campbell,friedlander}
studied soft constraints that allow incorporation of prior assumptions
into the MAXENT estimator making it effectively parametric. Here MAXENT
will be taken in its original meaning as providing non-parametric
estimators. 

This paper is organized as follows. Section \ref{sdata} recall the
tenets of the Bayesian decision theory and describes the data-generation
set-up.  Section \ref{sBayes} introduces and motivates the Bayesian
estimator and the regularized ML estimator. Section \ref{sMAXENT}
recalls the basic formulas of MAXENT, applies them to the studied
set-up, and discusses their symmetry features. Section \ref{sNum}
compares predictions of MAXENT with the regularized ML. We close in the
last section with discussing open problems. Appendix \ref{categorical}
shows how to apply MAXENT to categorical data. Appendix \ref{minimum}
presents our preliminary results on the affine symmetry of MAXENT 
estimators, and establishes
relations with the minimum entropy principle proposed in 
\citep{good_entropy,christ,minent,mdl}.

\section{Bayesian decision theory}
\label{sdata}

Consider a random quantity $Z$ with values $(z_1,...,z_n)$ and 
respective probabilities $q=(q_1,...,q_n)=(q(z_1),...,q(z_n))$.
We look at an i.i.d. sample of length $M$:
\BEA
\label{sampo}
\D=(Z_{1},...,Z_{M}),\qquad m=\{m_k\}_{k=1}^n, \qquad M\equiv\sum_{k=1}^nm_k,
\EEA
where $Z_u\in (z_1,...,z_n)$ ($u=1,...,M$), and 
$m_k$ is the number of appearances of $z_k$ in (\ref{sampo}). 
This sample will be an instance of our data, e.g. constraints of MAXENT
will be determined from it. The conditional probability of data $\D$
reads
\BEA
\label{pulti}
P(\D|q_1,...,q_n)=
P(m_1,...,m_n|q_1,...,q_n)=M!\prod_{k=1}^n \frac{q_k^{m_k}}{m_k!}.
\EEA

To check the performance of various inference methods, the probabilities 
$\hat q(\D)=\{\hat q_k(\D)\}_{k=1}^n$ inferred from (\ref{sampo})
are compared with true probabilities $q=\{q(z_k)\}_{k=1}^n$ via 
the KL-distance 
\BEA
\label{KL}
K[q,\hat q(\D)] = \sum_{k=1}^n q_k\ln\frac{q_k}{\hat q_k(\D)},
\EEA
where concrete forms of $\hat q(\D)$ are given below. 
The choice of distance (\ref{KL}) is motivated below, where we recall that it
implies the global optimality of the standard (posterior-mean) Bayesian
estimator. Another possible choice of distance is the squared
(symmetric) Hellinger distance: ${\rm dist}_{\rm H}[q,\hat q]\equiv
1-{\sum}_{k=1}^n\sqrt{ q_k\, \hat q_k}$. In our situation, it frequently
leads to the same qualitative results as (\ref{KL}). 

How to compare various estimators with each other, and decide on the
quality of a given estimator?  Bayesian decision theory comes to answer
this question; see chapter 11 of \citep{cox}. The theory assumes that
the probabilities of $(z_1,...,z_n)$ are generated from a known
probability density $P(q_1,...,q_n)$ that encapsulates the prior
information about the situation. Next it decides on the quality of an
estimator $\hat q(\D)$ via the average distance
\BEA
\label{bat}
\langle \overline{ K }\rangle =\int\prod_{k=1}^n\d q_k\, P(q_1,...,q_n)\,
\overline{K\,},\qquad
\overline{K\,}=\sum_{\D} P(\D|q) K[q,\hat q(\D)].
\EEA
where $\overline{K\,}$ is the average of (\ref{KL}) over samples
(\ref{sampo}) with fixed length $M$.  Sometimes the Bayesian decision
theory replaces the distance by the utility, loss {\it etc} 
In the Bayesian decision theory different loss (or decision) functions can be 
optimized based on the context of the problem \citep{cox}.
Note the difference between the proper Bayesian approach and the
Bayesian decision theory; cf.~chapters 10 and 11 in \citep{cox}. The
former employs the data for moving from the prior (\ref{di}) to the
posterior (\ref{eq:9}). It averages over the prior, e.g. when
calculating the posterior mean. The latter advises on choosing
estimators, whose form may or not may not depend on the prior; see below
for examples. The decision theory averages both over the data and over the
prior, as seen in (\ref{bat}).

For the prior density of $q=\{q_k\}_{k=1}^n$ we choose the Dirichlet density
(or a mixture of such densities as seen below) \citep{washington,schafer}:
\BEA
\label{di}
P(q_1,...,q_n\,;\,\alpha_1,...,\alpha_n)=\frac{\Gamma[\sum_{k=1}^n \alpha_k]}{\prod_{k=1}^n \Gamma[\alpha_k]   }
\prod_{k=1}^n q_k^{\alpha_k-1}\, \delta(\sum_{k=1}^nq_k-1),
\EEA
where $\Gamma[x]=\int_0^\infty\d y\, y^{x-1}\,e^{-y}$ is Euler's $\Gamma$-function and
delta-function $\delta(\sum_{k=1}^nq_k-1)$ ensures the normalization of probabilities.
Parameters $\alpha_k>0$ determine the prior weight of $q_k$ \citep{washington,schafer}:
\BEA
\label{di101}
\langle q_k\rangle\equiv \int_0^\infty\prod_{l=1}^n \d q_l\, q_k\, 
P(q_1,...,q_n\,;\,\alpha_1,...,\alpha_n)
=\frac{\alpha_k}{A}, \qquad A\equiv\sum_{k=1}^n \alpha_k,
\EEA
where the integration range goes over the simplex $0\leq q_k\leq 1$, $\forall k$, and $\sum_{k=1}^n q_k=1$.
Dirichlet density (\ref{di}) is unique in holding several desired features of non-informative
prior density over unknown probabilities; see
\citep{washington,schafer} for reviews. An important feature of density (\ref{di}) is that it is
conjugate to the multinomial conditional probability (\ref{pulti})
\begin{eqnarray}
  \label{eq:9}
P(q_1,...,q_n|m_1,...,m_n)= P(q_1,...,q_n\,;\,\alpha_1+m_1,...,\alpha_n+m_n).
\end{eqnarray}
Eq.~(\ref{eq:9}) is convenient when studying i.i.d.
samples (\ref{sampo}) of discrete random quantities. Here we assume that the prior
density is known exactly [see however (\ref{perturb})]. In practice, such a knowledge
need not be available. For example, it may be known that the prior density belongs to 
the Dirichlet family, but its hyper-parameters $\{\alpha_k\}_{k=1}^n$ are unknown and should be determined 
from the data, e.g. via empirical Bayes procedures; see \citep{washington,schafer,claesen,ran,berg} for reviews on
hyper-parameter estimation. 

\comment{
We mention one such feature that is employed below
in our calculations. 
The random variables with realizations $(q_1,...,q_n)$ are
conditionally independent, i.e. they are independent modulo the
constraint that they sum to $1$; see (\ref{di}). On the other hand, (\ref{di}) can be
sampled by normalizing independent positive random variables
$\{\zeta_k\}_{k=1}^{n}$ with probability density $\rho(\zeta)$: $\zeta_k\sim\rho(\zeta_k)$.
One defines after normalization:
$q_k=\frac{\zeta_k}{\sum_{l=1}^n \zeta_l}$ ($k=1,...,n$).
If we now assume that $\rho(\zeta_k)$ follows the $\Gamma$-distribution:
$\rho(\zeta_k)=\frac{\zeta^{\alpha_k-1}\,e^{-\zeta}}{\Gamma[\alpha_k]}$,
we revert to (\ref{di}). Now the $\Gamma$-distribution 
is the only density of $\zeta_k>0$ in (\ref{torro}) that ensures the above 
feature of conditional independence \citep{washington,schafer}. These two
independence features characterize (\ref{di}) uniquely
\citep{washington,schafer}.
}

\section{Bayesian and regularized maximum likelihood (ML) estimators}
\label{sBayes}

Starting from (\ref{bat}), we find the best estimator in terms of the minimal, 
average KL-distance: 
\BEA
\label{bato}
{\rm min}[\, \langle \overline{ K^{} }\rangle \,]
=\sum_\D P(\D)\, {\rm min}\,\left[\, \int\prod_{k=1}^n\d q_k\,P(q|\D)K[q,\hat q(\D)]
\right],
\EEA
where the minimization goes over inferred probabilities $\{\hat
q(\D)\}$, and where $P(q|\D)$ is recovered from $P(\D|q)$:
$P(\D)P(q|\D)=P(\D|q)P(q)$; cf.~(\ref{sampo}, \ref{pulti}).  The
equality in (\ref{bato}) follows from the fact that if $\hat q(\D)$
minimizes $\int\prod_{k=1}^n\d q_k\,P(q|\D)K[q,\hat q(\D)]$, then it
will minimize each term of the sum for every $\D$, and thus will
minimize the whole sum.  Then implementing the constraint $\sum_{k=1}^n
\hat q_k(\D)=1$ via a Lagrange multiplier, we get from (\ref{bato}):
\BEA
\label{gor}
{\rm argmin}\,\left[\, \int\prod_{k=1}^n\d q_k\,P(q|\D)\,K[q,\hat q(\D)]\,\right]
=\left\{ \int\prod_{k=1}^n\d q_k\,q_l\,P(q|\D)\right \}_{l=1}^n.
\EEA
We got in (\ref{gor}) the posterior average, because we employed the KL
distance $K[q,\hat q(\D)]$. The optimal estimator will be different upon
using another distance, e.g.  KL distance $K[\hat q(\D),q]$ of $\hat
q(\D)$ from $q$, or the Hellinger distance. Note that in the
proper Bayesian approach the posterior mean is simply postulated to be
an estimator, since it is just a characteristics of the posterior
distribution. In the present Bayesian decision approach the posterior
emerges from minimizing a specific ({\it viz.} KL) distance. If another
distance is used, the posterior mean is not anymore optimal.  

If the prior is a single Dirichlet density (\ref{di}) we get from (\ref{eq:9}, \ref{gor})
for the Bayesian estimator:
\BEA
\label{lid}
p(z_k)=\frac{m_k+\alpha_k}{M+A}.
\EEA 
The average KL-distance (\ref{bat}) for the estimator (\ref{lid}) reads from 
(\ref{eq:9}, \ref{pulti}) (denoting $\psi[x]\equiv\frac{\d}{\d x}\ln\Gamma[x]$):
\begin{align}
&\langle\overline{K[q,p]}\rangle=
\frac{1}{A}\sum_{k=1}^n\alpha_k\psi(1+\alpha_k)-\psi(1+A) +\ln(M+A) \nonumber\\
&-\frac{\Gamma[M+1]\,\Gamma[A]}{\Gamma[M+A+1]}\,
\sum_{k=1}^n \sum_{m=0}^M \frac{\Gamma[m+1+\alpha_k]\,\Gamma[M-m+A+\alpha_k]\,\ln(m+\alpha_k)}
{\Gamma[\alpha_k]\,\Gamma[A-\alpha_k]\,\Gamma[m+1]\,\Gamma[M-m+1]}.
\label{KL2}
\end{align}

If the prior density is given by mixture of Dirichlet densities with weights 
$\{\pi_a\}_{a=1}^L$:
\BEA
\label{mixture}
\sum_{a=1}^L\pi_a P(q_1,...,q_n\,;\,\alpha^{[a]}_1,...,\alpha^{[a]}_n),\qquad \sum_{a=1}^L\pi_a=1,
\EEA
then instead of (\ref{di101}) and (\ref{lid}) we have from (\ref{gor})
\BEA
\label{di1010}
&&\langle q_k\rangle
=\sum_{a=1}^L\pi_a\frac{\alpha^{[a]}_k}{A^{[a]}}, \qquad A^{[a]}\equiv\sum_{k=1}^n \alpha^{[a]}_k,\\
\label{lido}
&&p(z_k)=\frac{\sum_{a=1}^L\pi_a\,\, \Phi^{[a]}\, \,\frac{m_k+\alpha_k^{[a]} }{M+A^{[a]}} 
 } {\sum_{a=1}^L\pi_a\, \Phi^{[a]} },\qquad
\Phi^{[a]}
\equiv\frac{\Gamma[A^{[a]}]}{\Gamma[M+A^{[a]}]}\,\prod_{k=1}^n\frac{\Gamma[m_k+\alpha_k^{[a]} ]}{\Gamma[\alpha_k^{[a]}]}.
\EEA
For a mixture prior density, the Bayesian estimator (\ref{lido}) depends on {\it all} numbers 
$\{m_k; \alpha^{[1]}_k,...,\alpha^{[L]}_k\}$ not just on $m_k$. Below we illustrate
that not knowing precisely details of the prior mixture can lead to serious losses
when applying Bayesian estimators. 

It is interesting (both conceptually and practically) to have a simple
estimator, where the dependence on the prior is reduced to a single
parameter. A good candidate is the regularized maximum likelihood (ML)
estimator (see \citep{ml_review} for a review):
\BEA
\label{lidd}
p_{\rm ML}(z_k)\equiv\frac{m_k+b}{M+nb}=\lambda \frac{m_k}{M} + (1-\lambda) \frac{1}{n}, 
\quad \lambda = \frac{M}{M + nb}, \qquad 
b\geq 0,  \quad 0<\lambda<1,
\EEA 
where the regularizer $b$ (or $\lambda$) takes care of the fact that for
a finite sample (\ref{sampo}) not all values $z_k$ had a chance to
appear (i.e. $m_k=0$ for them). Then (\ref{lidd}) avoids to claim a zero
probability due to $b>0$. Eq.~(\ref{lidd}) is a shrinkage estimator,
where the proper ML estimator $\frac{m_k}{M}$ is shrunk towards uniform
distribution $\frac{1}{n}$ by the shrinkage factor $\lambda $.  The
proper ML estimator $p_{\rm ML}(z_k)|_{b=0}$ will be shown to be a
meaningless estimator for not very long samples (\ref{sampo}) producing
results that are worse than $\{q(z_k)=\frac{1}{n}\}_{k=1}^n$. Moreover,
for such samples the correct choice of $b$ (based on the prior
information) is crucial, i.e. (\ref{lidd}) is generally a parametric 
estimator. The estimator (\ref{lidd}) recovers true
probabilities for $M\to \infty$ \citep{cox}, where $n$ and $b$ are
fixed, hence $\lambda\to 1$ in (\ref{lidd}). 

For the optimal estimator (\ref{lidd}), the value of $b$ is found by
minimizing the average KL-distance (\ref{bat}). When the prior is given
by a Dirichlet density (\ref{di}), the average KL-distance amounts to
(\ref{KL2}), where we need to replace $\ln(M+A)\to \ln(M+nb)$ and
$\ln(m+\alpha_k)\to \ln(m+b\,)$. Now (\ref{gor}, \ref{lid}) imply that
for a homogeneous Dirichlet prior, i.e. for (\ref{di}) with
$\alpha_k=\alpha$, we have $b_{\rm opt}=\alpha$ for the optimal value of
$b$, i.e. the regularized ML estimator coincides with the Bayesian
estimator: $p_{\rm ML}(z_k)=p(z_k)$.  This does not anymore hold for the
mixture of Dirichlet prior densities. 

\section{The maximum entropy (MAXENT) method }
\label{sMAXENT}

MAXENT infers probabilities from maximizing the 
Boltzmann-Gibbs-Shannon entropy
\BEA
\label{ento} S[q]= -\sum_{k=1}^n q(z_k)\ln q(z_k), 
\EEA
under constraints taken from the sample (\ref{sampo}). The rationale of
maximizing (\ref{ento}) is that a larger $S$ means a smaller bias (or
information) according to several axiomatic schemes
\citep{jaynes,chakra,baez,cover,topsoe,shore,paris,balian,rmp}.  Note
that physical applications of MAXENT operate with constraints that are
known precisely, e.g. the mean energy constraint is deduced from the
corresponding conservation law \citep{jaynes,balian,rmp}. Such
situations are rare in statistics and machine learning. Hence we need to
understand which constraints are to be taken from the noisy data.

First we can apply no constraint and maximize the entropy:
\BEA
q^{[0]}(z_k)={1}/{n}.
\label{grundig}
\EEA
The calculation of the average distance is straightforward from 
(\ref{bat}, \ref{KL2}, \ref{grundig}) both for a single Dirichlet
prior and a mixture of such priors. We examplify the 
single Dirichlet case (\ref{di}):
\BEA
\langle\overline{K[q,q^{[0]}]}\rangle= \sum_{k=1}^n \langle q_k\ln{q_k}\rangle +\ln n=
\frac{1}{A}\sum_{k=1}^n\alpha_k\psi(1+\alpha_k)-\psi(1+A)+\ln n.
\label{bruno}
\EEA
Now $\langle\overline{K[q,q^{[0]}]}\rangle$ plays an important role:
once (\ref{grundig}) is completely data-independent and simply
reproduced the prior expectation on the unbiased probabilities,
estimators that provide the average KL-distance larger than
(\ref{bruno}) are meaningless; see below for examples. 

Next, we employ the empiric mean of (\ref{sampo}) as a constraint for the 
expected value of $Z$:
\BEA
\label{first}
\mu_1=\frac{1}{M}\sum_{u=1}^M Z_{u}=\frac{1}{M}\sum_{k=1}^n z_{k}m_{k} =\sum_{k=1}^n q_kz_k.
\EEA
Maximizing (\ref{ento}) under constraint (\ref{first}) via the Lagrange 
method leads to the famous Gibbs formula \citep{jaynes,balian,rmp}:
\BEA
\label{s1}
q^{[1]}(z_k)=\frac{e^{-\beta z_k}}{\sum_{l=1}^n e^{-\beta z_l} },
\EEA
where the Lagrange multiplier $\beta$ is found from (\ref{first}). 
Appendix presents an example of applying (\ref{s1}) to real data. 
We order the values of $Z$ as $z_1<...<z_n$, and note a
specific feature of (\ref{s1}): depending on the sign of $\beta$, we either get 
\BEA
\label{vok}
q^{[1]}(z_1)\leq ... \leq q^{[1]}(z_n)\quad {\rm or}\quad q^{[1]}(z_1)\geq
... \geq q^{[1]}(z_n). 
\EEA
One can try to acquire further information from sample (\ref{sampo}) by looking at the second empiric moment:
\BEA
\label{second}
\mu_2=\frac{1}{M}\sum_{u=1}^M Z^2_{u}=\frac{1}{M}\sum_{k=1}^n z^2_{k}m_{k} =\sum_{k=1}^n q_kz^2_k.
\EEA
Now we maximize (\ref{ento}) under two constraints (\ref{first}) and (\ref{second}):
\BEA
\label{s2}
q^{[1+2]}(z_k)=\frac{e^{-\beta_1 z_k-\beta_2 z_k^2}}{\sum_{l=1}^n e^{-\beta_1 z_l-\beta_2 z_l^2} },
\EEA
where Lagrange multipliers $\beta_1$ and $\beta_2$ are found from
solving both (\ref{first}) and (\ref{second}). Eqs.~(\ref{s1},
\ref{s2}) make obvious how to involve other (fractional)
moments. The maximizations of (\ref{ento}) lead to unique results, because (\ref{ento}) is a
concave function of $\{p_k\}_{k=1}^n$, while the moment constraints are
linear. 

Let the values $(z_1,...,z_n)$ of $Z$ be subject to affine transformation
\begin{gather} 
\label{nomo} 
\widetilde{z}_k={\cal F}(z_k), \quad {\cal
F}(z)=g z+h, \quad k=1,...,n.  
\end{gather}
Hence as a result of transformation (\ref{nomo}):
$\mu_1\to\widetilde{\mu}_1=g\mu_1+h$ and
$\mu_2\to\widetilde{\mu}_2=g^2\mu_2+h^2+2gh\mu_1$; see (\ref{first}, \ref{second}). These
relations show that (\ref{nomo}) leaves
the inferred probabilities (\ref{s1}, \ref{s2}) invariant, because
the resulting set of equations for the unknowns in (\ref{s1}, \ref{s2})
are identical for both the original $(z_1,..., z_n)$ and transformed values (\ref{nomo}).
Likewise, involving first
$p$ moments $\mu_1,...,\mu_p$ produces affine-invariant
probabilities. Note that involving only (\ref{second}) [without involving
(\ref{first})] will lead to the invariance of the probabilities with
respect to a limited affine-symmetry, where $h=0$ in (\ref{nomo}).
Another example of limited affine symmetry is involving the fractional
moment $\sum_{k=1}^n q_k\sqrt{z_k}$ (for $z_k\geq 0$ and instead of
(\ref{first}, \ref{second})). Then the probabilities
$q^{[1/2]}(z_k)\propto e^{-\beta_{1/2} \sqrt{z}_k}$ will stay intact
only under $h=0$ and $g>0$ in (\ref{nomo}). Note in this context that
the ML estimator (\ref{lidd}) is invariant with respect (\ref{nomo})
with an arbitrary bijective ${\cal F}$, which keeps the values of
$\widetilde{z}_k$ different. 

The symmetry features of various estimators are clearly important,
though we so far have no analytical results that would relate them to
the estimation quality quantified by (\ref{bat}). But we noted from
numerical comparison of MAXENT estimators based on various constraints,
that estimators with the largest affine symmetry, i.e. (\ref{nomo}) with
arbitrary $g$ and $h$, tend to be better in terms of the average
KL-distance (\ref{bat}). Intuitively, higher (affine) symmetry should be
related to higher susceptibility with respect to noises; see Appendix 
\ref{minimum} for further results.

\comment{Note that MAXENT has applicability limits related to small values of
$M$: for samples that consist of only $z_1$'s or only $z_n$'s (i.e.
samples with $m_1=M$ or $m_n=M$) we get $q^{[1]}_k=\delta_{k1}$ and
$q^{[1]}_k=\delta_{kn}$, respectively, where $\delta_{ik}$ is the
Kronecker delta. This implies that the KL-distance
$K[q,q^{[1]}]$ is infinite for those samples, and hence
also $\overline{K[q,q^{[1]}]\,}=\infty$. However, for 
\BEA
\label{ina}
n^{-M}\ll 1, 
\EEA
the MAXENT can be regularized such that the infinities are avoided. Now $n^{-M}$ 
is a rough estimate for the probability of the above dangerous
samlpes and hence (\ref{ina}) means that their appearance will not be noticed.  
}

\section{Numerical results}
\label{sNum}

\subsection{A single Dirichlet density}

Recall that maximization of entropy (\ref{ento}) can
be applied if there is no prior information that distinguishes one 
probability from another. If such information is present, MAXENT is generalized to
the minimum relative entropy method \citep{shore}. We shall not study
this generalization here. Hence to ensure applicability of MAXENT, we
always choose prior densities such that $\langle q_k\rangle=\frac{1}{n}$;
i.e. all $n$ values are equally likely to be generated, on average.  
As seen from (\ref{di101}), for
a single prior Dirichlet density (\ref{di}) condition $\langle q_k\rangle=\frac{1}{n}$ 
implies:
\BEA
\label{diro}
\alpha_k=\alpha,\qquad k=1,...,n,
\EEA
Now recall from (\ref{lidd}, \ref{lid})
that under (\ref{diro}) the Bayesian and the regularized ML coincide.
I.e. we conclude  that the
regularized ML is a better estimator than MAXENT (under any constraint).

Though $\langle q_k\rangle=\frac{1}{n}$ does not depend on $\alpha$, the
most probable values $\widetilde{q}_k$ of $q_k$ do depend on the
magnitude of $\alpha$.  Finding $\widetilde{q}_k$ from (\ref{di},
\ref{diro}) amounts to maximizing ${\cal L}(q)=(\alpha-1)\sum_{k=1}^n\ln
q_k +\gamma \sum_{k=1}^nq_k$, where the Lagrange multiplier $\gamma$
ensures $\sum_{k=1}^nq_k=1$. For $\alpha>1$, ${\cal L}(q)$ is a concave
function of $q$, and its global maximum is found after differentiating
it. Hence $\widetilde{q}_k$ holds
\BEA
\label{bruce22}
\widetilde{q}_k= 1/n \quad {\rm for} \quad \alpha>1, \quad k=1,...,n.
\EEA
For $\alpha<1$, ${\cal L}(q)$ is a convex function, it does not have
local maxima with $q_k>0$ ($k=1,...,n$).  Its maxima are located at
points, where $q_k=0$ for certain $k$. Repeating this argument, we see
that the maxima of ${\cal L}(q)$ are at those points, where a possibly 
large number of $q_k$ are zero:
\BEA
\label{bruce11}
\widetilde{q}_k= 0\quad {\rm or}\quad \widetilde{q}_k= 1,
\quad {\rm for} \quad \alpha<1, \quad k=1,...,n,
\EEA
which means deterministic probabilities. Eq.~(\ref{bruce11})  
is consistent with $\langle q_k\rangle=1/n$, because 
there are $n$ equivalent most probable values. 
\comment{
Note that the Dirichlet distribution under (\ref{diro}) and
$\alpha\to 0$ is sometimes regarded as non-informative in the sense of
Haldane; see \cite{jaeger} for a recent review.  To our opinion, the
very fact that it has deterministic most probable values does constitute
a form of prior information. }

Let us start with the regime $\alpha>1$; cf.~(\ref{bruce22}).
Table \ref{tab0} compares predictions of (\ref{lidd}) with those of
MAXENT solutions (\ref{s1}) and (\ref{s2}) for the Dirichlet prior
(\ref{di}) holding (\ref{diro}) with $\alpha=2$. It is seen that 
MAXENT is meaningless, because the trivial estimator (\ref{grundig})
provides a smaller average KL-distance; cf.~(\ref{bruno}). For the Bayesian
estimator even $M=1$ leads to a meaningfull prediction; e.g.  for
parameters of Table \ref{tab0} we have: $\langle \overline{K^{}}_{\rm
Bayes}\rangle|_{M=1}=0.224<0.225$. 

\begin{table}
\centering
\tabcolsep0.058in \arrayrulewidth0.7pt
\renewcommand{\arraystretch}{1.3}
\caption{ 
For $n=60$ and $z_k=k$ ($k=1,...,n$) we show the average KL-distance (\ref{bat}) for
various estimators. The full affine symmetry (\ref{nomo}) holds for all shown probabilities.
$M$ is the length of sample (\ref{sampo}). The initial prior
Dirichlet density (\ref{di}) holds (\ref{diro}) with $\alpha_k=2$.
Eq.~(\ref{bruno}) equals $\langle\overline{K[q,q^{[0]}]}\rangle=0.225$, i.e. values of
the average KL-distance larger than $0.225$ are \emph{meaningless}.
$\langle \overline{K}_{\rm Bayes}\rangle$ is the averaged KL-distance for the Bayes estimator
(\ref{lid}) that for this case coincides with the optimally regularized ML estimator.  
$\langle\overline{K}_{1}\rangle$ and $\langle\overline{K}_{1+2}\rangle$
are defined (resp.) via (\ref{s1}, \ref{first}) and (\ref{s2}, \ref{second}).  
The averages are found numerically (applies to all Tables): first we generate $10^2$ instances of $\{q_k\}_{k=1}^n$ from 
the Dirichlet density, and then for each instance we generate $10^2$ samples (\ref{sampo}). 
Such parameters lead to 3-digit precision, as reported. 
}
\begin{tabular}{|l||c|c|c|}
\hline 
$M$ & 
$\langle \overline{K}_{\rm Bayes}\rangle$ & 
$\langle\overline{K}_{1}  \rangle$ & 
$\langle\overline{K}_{1+2}\rangle$  \\ \hline 
35  & 0.177 & 0.236      & 0.247    \\ 
25  & 0.188 & 0.240      & 0.260     \\ 
15  & 0.202 & 0.259      & 0.301     \\ \hline
\end{tabular}
\label{tab0} 
\end{table}

\begin{table}[h!]
\centering
\tabcolsep0.058in \arrayrulewidth0.7pt
\renewcommand{\arraystretch}{1.3}
\caption{ 
The same as in Table~\ref{tab0}, but for $\alpha_k=\alpha=0.1$ in (\ref{diro}).  
Eq.~(\ref{bruno}) gives $\langle\overline{K[q,q^{[0]}]}\rangle=1.798$, i.e. values of
the average KL-distance larger than $1.798$ are {meaningless}.
}
\begin{tabular}{|l||c|c|c|}
\hline
$M$ & 
$\langle \overline{K^{}}_{\rm Bayes}\rangle$ & 
$\langle\overline{K^{}}_{1}\rangle$ & 
$\langle\overline{K^{}}_{1+2}\rangle$  \\ \hline 
55  & 0.233 & 1.756      & 1.685    \\ 
45  & 0.276 & 1.700      & 1.643     \\
35  & 0.338 & 1.723      & 1.680     \\
25  & 0.428 & 1.753      & 1.717     \\
15  & 0.606 & 1.770      & 2.164     \\
11  & 0.730 & 1.762      & 4.946     \\
9   & 0.818 & 1.774      & 11.63     \\
7   & 0.916 & 1.848      & 32.24     \\
\hline
\end{tabular}
\label{tab01} 
\end{table}

\begin{table}[h!]
\centering
\tabcolsep0.058in \arrayrulewidth0.7pt
\renewcommand{\arraystretch}{1.3}
\caption{ The same as in Table~\ref{tab0}, but 
the initial prior density is a Dirichlet 
mixture given by (\ref{mixture}, \ref{so}) with $\alpha_0=0.3$ and $\epsilon=1.1$. The average KL-distance
$\langle\overline{K[q,q^{[0]}]}\rangle$ for the trivial estimator (\ref{grundig})
equals $0.212$, i.e. values of the average KL-distance larger
than $0.212$ are \emph{meaningless}; cf.~(\ref{bruno}). $\langle \overline{K}_{\rm Bayes}\rangle$ and
$\langle \overline{\bf K}_{\rm Bayes}\rangle$ refer to (\ref{lido}) and
(\ref{perturb}), respectively.
$\langle \overline{K}_{\rm ML}\rangle_{b=b_{\rm opt}}$ and $\langle \overline{K}_{\rm ML}\rangle_{b=1}$
refer to regularized ML estimator (\ref{lidd}) under $b=1$ and the optimal value of $b$ found from numerically minimizing 
$\langle \overline{K}_{\rm ML}\rangle$. The optimal value $b_{\rm opt}$ of $b$ changes from $2.46$ for $M=35$
to $2.65$ for $M=1$. We also report the value of $\langle \overline{K}_{\rm ML}\rangle_{b=1}$ with a sensible
value of $b$ to confirm that if $b$ is not chosen properly, 
then the corresponding (regularized) ML estimator (\ref{lidd}) is meaningless. 
$\langle\overline{K}_{1}\rangle$ is defined via (\ref{bat}, \ref{s1}). 
$\langle\overline{K}_{1+2}\rangle$ is not
shown, since $\langle\overline{K}_{1+2}\rangle> \langle\overline{K}_{1}\rangle$
for $35\geq M\geq 1$. We do not show $\langle\overline{K}_{1}\rangle|_{M=1}$, since it is larger than 
the average KL-distance for all other estimators.
}
\begin{tabular}{|l||cc|cc|c|}
\hline
$M$ & 
$\langle \overline{K}_{\rm Bayes}\rangle$ & 
$\langle \overline{\bf K}_{\rm Bayes}\rangle$ & 
$\langle \overline{K}_{\rm ML}\rangle_{b=b_{\rm opt}}$ & 
$\langle \overline{K}_{\rm ML}\rangle_{b=1}$ & 
$\langle\overline{K}_{1} \rangle$  \\ \hline 
35  & 0.014 & 0.206  & 0.180 & 0.204  & 0.048    \\ 
25  & 0.015 & 0.207  & 0.188 & 0.210  & 0.053    \\ 
15  & 0.017 & 0.209  & 0.197 & 0.214  & 0.065   \\ 
11  & 0.022 & 0.209  & 0.201 & 0.215  & 0.077    \\ 
7   & 0.035 & 0.209  & 0.205 & 0.215  & 0.105    \\ 
5   & 0.052 & 0.210  & 0.207 & 0.214  & 0.141    \\  \hline
3   & 0.083 & 0.210  & 0.209 & 0.213  & 0.268    \\ 
1   & 0.150 & 0.211  & 0.211 & 0.212  & ---       \\ \hline
\end{tabular}
\label{tab1} 
\end{table}

\begin{table}
\centering
\tabcolsep0.058in \arrayrulewidth0.7pt
\renewcommand{\arraystretch}{1.3}
\caption{
The same as in Table~\ref{tab1}, but for different values of $M$. Here
$\langle\overline{K}_{1+2}\rangle$ refers to MAXENT estimator (\ref{s2}) with constraints (\ref{first}, \ref{second}).
The Bayesian estimator is found from
(\ref{lido}, \ref{so}). For this range of sufficiently large $M$ the
MAXENT estimator (\ref{s2}) performs better than (\ref{s2}):
$\langle\overline{K}_{1+2}\rangle <\langle\overline{K}_{1}\rangle< \langle \overline{K}_{\rm ML}\rangle_{b=b_{\rm opt}}$. 
}
\begin{tabular}{|l||c|cc|cc|}
\hline
$M$ & 
$\langle \overline{K}_{\rm Bayes}\rangle$ & 
$\langle \overline{K}_{\rm ML}\rangle_{b=b_{\rm opt}}$ & 
$\langle \overline{K}_{\rm ML}\rangle_{b=1}$ & 
$\langle\overline{K}_{1} \rangle$ & 
$\langle\overline{K}_{1+2} \rangle$ \\ \hline 
45  & 0.015 & 0.172  & 0.196 & 0.045  & 0.042    \\ 
65  & 0.014 & 0.157  & 0.180 & 0.042  & 0.035    \\ 
85  & 0.014 & 0.145  & 0.164 & 0.040  & 0.031    \\ 
241 & 0.013 & 0.087  & 0.091 & 0.038  & 0.024     \\ \hline
\end{tabular}
\label{tab2} 
\end{table}

The above conclusion holds more generally (as we 
checked numerically): for the homogeneous Dirichlet
prior (\ref{diro}) with $\alpha\geq 1$, MAXENT estimators (\ref{s1},
\ref{first}) and (\ref{s2}, \ref{first}, \ref{second}) are meaningless
at least in the sparse data regime $M<n$. This puts a serious limitation
on the validity of MAXENT. 

The situation changes for sufficiently small values of $\alpha$ in the
regime (\ref{bruce11}); see Table~\ref{tab01} for $\alpha=0.1$. Here the
MAXENT estimators are meaningful provided that the sample length $M$ is
sufficiently large (but still in the sparse data regime $M<n$): (\ref{s1},
\ref{first}) is meaningful for $M\geq 9$ ($M<n=60$), while the estimator
(\ref{s2}, \ref{first}, \ref{second}) is meaningful for $M\geq 25$; see
Table~\ref{tab01}.  Though predictions of MAXENT are still far from
those of the Bayesian estimator, we should recall that the latter
estimator is parametric, i.e. it depends on the prior (via the parameter
$\alpha$) in contrast to MAXENT estimators. Table~\ref{tab01}
demonstrates the overfitting phenomenon: for $9\leq M\leq 15$ the MAXENT
estimator (\ref{s1}, \ref{first}) is meaningful, but adding the second
constraint makes the MAXENT estimator (\ref{s2}, \ref{first},
\ref{second}) not meaningful.  The situation is worsened since
(\ref{second}) is again estimated from the noisy data and gathers more
noise than information. This overfitting disappears for larger values of
$M$, i.e. $M\geq 25$, as Table \ref{tab01} demonstrates. Now adding the
second constraint (\ref{second}) is beneficial.

\subsection{Mixture of Dirichlet densities}

For modeling more complex types of prior information about the unknown
probabilities $\{q_k\}_{k=1}^n$, we shall assume that the prior density
is a mixture of two Dirichlet densities; see (\ref{mixture}).  Relations
$\langle q_k\rangle=\frac{1}{n}$ ($k=1,...,n$) will be still kept, since
they are necessary for applying MAXENT.  Now we assume that that there
are (prior) {\it conditional} rank correlation between the values
$(z_1,...,z_n)$ of $Z$, ordered as $(z_1<...<z_n)$, and its probabilities
$(q_1,...,q_n)$. For one component of the mixture, the probabilities
$(q_1,...,q_n)$ prefer to be ordered as in $(q_1<...<q_n)$. For another
component they tend to be ordered in the opposite way $(q_1>...>q_n)$.
This type of prior knowledge can be modeled via a mixture
(\ref{mixture}) of two Dirichlet priors with $L=2$,
$\pi_1=\pi_2=\frac{1}{2}$, and
\BEA
\label{ki}
&&\alpha_1^{[1]}<...<\alpha_n^{[1]}, \qquad 
\alpha_1^{[2]}>...>\alpha_n^{[2]},\\
&&\frac{\alpha_k^{[1]}-\alpha_l^{[1]}}{A^{[1]}}=\frac{\alpha_l^{[2]}-\alpha_k^{[2]}}{A^{[2]}}, \quad {\rm for~any}\quad
k,l=1,...,n,
\label{ai}
\EEA
where (\ref{ai}) ensures the needed $\langle q_k\rangle=\frac{1}{n}$, as seen from (\ref{di1010}).
A simple case that leads to (\ref{ki}, \ref{ai}) is
\BEA
\label{so}
L=2, \quad \pi_1=\pi_2=\frac{1}{2},\quad
\alpha_k^{[1]}=\alpha_0+\epsilon (k-1),\quad \alpha_k^{[2]}=\alpha_0+\epsilon (n-k), \qquad k=1,...,n,
\EEA
where $A^{[1]}=A^{[2]}=\alpha_0 n+\frac{\epsilon n(n-1)}{2}$. 
Recall that for a mixture of Dirichlet densities the Bayes 
estimator (\ref{lido}) and the optimally regularized ML estimator 
(\ref{lidd}) are different. 

For numerical illustration we choose
$\{z_k=k\}_{k=1}^n$. Prior probability densities generated via (\ref{so}) will be
employed with $z_1<...<z_n$. 
Now Tables \ref{tab1} and \ref{tab2} show that for
$M\geq 5$ the MAXENT estimator (\ref{s1}, \ref{first}) is clearly better
than the {\it optimally} regularized ML estimator (\ref{lidd}):
\begin{align}
\label{burkina}
\langle\overline{K}_{1}\rangle<\langle \overline{K}_{\rm ML}\rangle_{b=b_{\rm opt}},
\end{align}
where the optimal value of $b$ is found from minimizing the averaged KL-distance (\ref{bat}). 
Moreover, for $M\geq 7$, we see that $\langle\overline{K}_{1}\rangle$ is closer to the optimal
$\langle \overline{K}_{\rm Bayes}\rangle$ than to $\langle \overline{K}_{\rm ML}\rangle_{b=b_{\rm opt}}$.
Note that such threshold values for $M$ do depend on the assumed prior
density and on $n$. 

For $M\to\infty$ the performance of the optimally regularized ML estimator
(\ref{lidd}) (for a fixed $b\sim 1$) will be better than MAXENT with any
finite number of constraints, since the regularized ML converges to the
true probabilities for $M\to\infty$ \citep{cox}, while MAXENT 
does not. But as Table \ref{tab2} shows, MAXENT with
constraints (\ref{first}) or (\ref{first})+(\ref{second}) still performs
better than the optimally regularized ML even for $M$ as large as 241
(for $n=60$). 

Table \ref{tab1} shows that MAXENT with two constraints (\ref{first},
\ref{second}) performs worse than the method under the single constraint
(\ref{first}) although the affine invariance (\ref{nomo}) of
probabilities holds.  This {\it overfitting} situation changes for
larger values of $M$, i.e. $M\geq 45$, as Table \ref{tab2} demonstrates. 

To stress the relevance of rank correlations, we note that the advantage (\ref{burkina}) of MAXENT
closely relates to the agreement between
(\ref{ki}) and the ordering $(z_1<...<z_n)$ of $Z$. If the vector
$(z_1<...<z_n)$ is randomly permuted, and employed for values of $Z$,
predictions of MAXENT become meaningless even for rather large values of $M>n$. 
\comment{To illustrate this point, let us take parameters of Table \ref{tab1}, but
apply a random permutation to values $\{z_k=k\}_{k=1}^{n=60}$. Taking this
new vector for values of $Z$, we noted that even for $M=85>n=60$ predictions of 
MAXENT under constraint (\ref{first}) is meaningless, since the average KL
distance amounts to $0.215$, which is larger than the prediction $0.212$ of
(\ref{bruno}). For $M=105$ this prediction $0.210$ is not meaningless in 
this sense, but rather poor by itself. It does not substantially improve even for
$M=500$ (for $n=60$), where it amounts to $0.207$. Though such predictions 
are not meaningless, they are practically useless. }

Recall that both the Bayesian (\ref{lido}) and the regularized ML
estimator (\ref{lidd}) are parametric, i.e. the very their form depends
on the prior, which is frequently not available in practice. Hence we
need to understand how strong is dependence. Let us assume that one has
to employ a Bayesian estimator without knowing the full form of the
equal-weight mixture (\ref{so}). Instead one knows the average values of
$\alpha_k=\frac{1}{2}[\alpha_0+\epsilon (k-1)]+
\frac{1}{2}[\alpha_0+\epsilon (n-k)]$ from (\ref{so}), prescribes them
to a single Dirichlet prior (\ref{di}, \ref{diro}) and builds up from
(\ref{lid}) an estimator
\BEA
\label{perturb}
{\bf p}(z_k)=\frac{m_k+\alpha_0+\epsilon (n-1)/2}{M+n\alpha_0+\epsilon n(n-1)/2}.
\EEA
The performance of this perturbed Bayesian estimator
deteriorates and gets worse than that of the MAXENT solution:
$\langle\overline{K}_{1}\rangle<\langle \overline{\bf K}_{\rm
Bayes}\rangle$; see Table \ref{tab1}. Likewise, the choice of $b$ in the
regularized ML estimator (\ref{lidd}) is important. If just some
reasonable value is chosen instead of the optimal one, e.g. $b=1$ instead
of $b\sim 2.5$ in Table \ref{tab1}, then the ML estimator can turn
meaningless; see Table \ref{tab1} for $M\leq 15$. In this context, we
emphasize that the Bayes estimator and the optimally regularized ML estimator are
never meaningless even for $M=1$. For parameters of Table \ref{tab1}, the MAXENT estimator 
(\ref{s1}) becomes meaningless for $M\leq 3$. 

\section{Summary and Discussion}

The maximum entropy (MAXENT) method provides non-parametric estimators
for inferring unknown probabilities \citep{volume}.  MAXENT is widely
applied both in statistical physics and probabilistic inference.
However, its physical applications are mostly data-free and are based on
additional principles (e.g. conservation laws \citep{balian,rmp}) that
are normally absent in statistics and machine learning. Hence we needed
a systematic approach towards understanding the validity limits of
MAXENT as an inference tool. 

Here we presented a Bayesian decision theory approach that allows to
determine on whether MAXENT is applicable at all, i.e. whether it is
better than a random guess. It also allows to compare different
estimators with each other (e.g. to compare MAXENT with the regularized
maximum likelihood), and study the relevance of various constraints
employed in MAXENT. 

Our results are summarized as follows. MAXENT does apply to a sparse
data, but demands specific prior information. Here sparse
means $M<n$, i.e. the sample length $M$ is smaller than the number of
probabilities $n$ to be inferred. We explored two different scenarios of
such prior information. First, the unknown probabilities generated by
homogeneous Dirichlet density (\ref{diro}) are most probably
deterministic. Second, there are prior rank correlations between the
random quantity and its probabilities. This seems to be the simplest
prior information that makes MAXENT applicable and superior over the
optimally regularized maximum-likelihood estimator.  Our approach is
capable of describing several phenomena that are relevant for applying
and understanding estimators: overfitting (i.e. adding more noisy
constraints leads to poorer inference), instability of optimal Bayesian
parametric estimators with respect to variation of prior details,
inapplicability of non-parametric MAXENT estimators to very short samples {\it
etc}. 

Several important problems were uncovered by this study and should be
addressed in future. First of all, this concerns the applicability of
MAXENT to a categorical data, where the values
$(z_1,...,z_n)$ of the random variable $Z$ in sample (\ref{sampo}) are
not numerical, but instead refer to certain distinguishable categories. The
major difference between maximum likelihood and MAXENT estimator is that
the former freely applies to categorical data. In contrast, MAXENT does
depend on the concrete numerical implementation (i.e. {\it encoding}) of
data, though this dependence is somewhat weakened by the affine symmetry
(\ref{nomo}). Thus an open problem demands considering various encoding
schemes in view of their applicability to MAXENT estimators. (In this
paper we in fact assumed the simplest encoding via natural numbers; see
Tables.) Appendix \ref{categorical} reports preliminary results in this direction along with 
a real data example. 
The second open problem relates to the influence of affine
symmetries on the performance of various MAXENT estimators. We observed
numerically that the constraints which produce affine-invariant
probabilities produce better estimators; see after (\ref{nomo}). 
Preliminary results along this direction are given in Appendix 
\ref{minimum}, where we also show relations of our results with
the minimum entropy principle proposed in 
\citep{good_entropy,christ,minent,mdl} for contraint selection.

\acknowledgments
We thank Roger Balian for useful discussions.

AEA and NHM were supported by SCS of Armenia, grants No. 18RF-015 and No. 18T-1C090.


\bibliography{maxentropy_arxiv}
\bibliographystyle{apsK}

\appendix
\section{MAXENT applies to categorical data }
\label{categorical}

The MAXENT method can be applied to any multinomial data (\ref{sampo}),
provided that numeric values $(z_1,...,z_n)$ of the random quantity $Z$
are given. MAXENT estimators depend on $(z_1,...,z_n)$ modulo the affine
symmetry (\ref{nomo}). This creates a problem in applying MAXENT to
categorical data, since for MAXENT one now needs a specific
encoding of categorical $Z$ into a numeric representation $(z_1,...,z_n)$
of categories. Recall that there is a degree of arbitrariness in choosing the
regularizer in maximum likelihood (ML) estimator, or in choosing prior
parameters for Bayesian inference. Here the arbitrariness lies in
different encodings. In practice, the proper encoding of categories
arises in any problem dealing with categorical data. If categories are
ordinal (e.g. military ranks, education levels), then one can use
$z_1<z_2<...< z_n$ encoding.  However, for nominal categories (e.g.
ethnicity, preference, disease) there is no such ordering. 

Let us illustrate the MAXENT method with the simple data from
pre-election presidential polling conducted in 1988, where out of $M =
1447$ voters $m_1 = 727$ preferred Bush, $m_2 = 583$ preferred Dukakis,
and $m_3 = 137$ preferred other candidates or had no preference. The
data together with its Bayesian analysis is taken from \citep{gelman}. 

Here our random variable $Z$ is voter's preference with three outcomes
$(z_1, z_2, z_3) = $ ('Bush', 'Dukakis', 'Other') and unknowns $(q_1,
q_2, q_3)$, which just represent the fractions of the population with
each preference. The goal here is to estimate $q_1 - q_2$, i.e. whether
Bush has more support than Dukakis. One can assume the Dirichlet
noninformative prior distribution for $(q_1, q_2, q_3)$ with parameters
$\alpha_1 = \alpha_2 = \alpha_3 = 1$, compute the posterior means of
$q_1$ and $q_2$ $(\hat q_1, \hat q_2)$ and take the difference
\citep{gelman}. The results show that Bush has more support: $\hat q_1 >
\hat q_2$.

Since the data is purely categorical, we shall apply the frequency
encoding for MAXENT: each category is represented with its frequency in the data
set, e.g. in this example $(z_1,z_2,z_3) = (0.502,0.403,0.095)$.  Now
empiric mean is equal to $0.42$ and the maximizing solutions of
(\ref{ento}) with $\sum_{k=1}^3 q_k z_k = 0.42$ are $(q^{[1]}(z_1),
q^{[1]}(z_2), q^{[1]}(z_3)) = (0.535,0.36,0.105)$. Thus, also the MAXENT result shows more support for Bush.

The more detailed data to the same problem from \citep{gelman} is displayed in Table~\ref{tab10}, where $M =
1447$ voters were stratified into $16$ regions. The Proportion column
shows the proportions $M_i/M (i=1,...,16)$ of voters registered in each
region, and $(m_{1i}/M_i, m_{2i}/M_i, m_{3i}/M_i)$ in each row are
proportions of voters preferring Bush, Dukakis, and others/no-preference
among those who vote in the corresponding region.  As in the previous
example, one can assume the Dirichlet prior distributions for $(q_{1i},
q_{2i}, q_{3i})$ with $\alpha_1 = \alpha_2 = \alpha_3 = 1$, this time
for each region separately and compute the posterior means of $q_{1i}$
and $q_{2i}$ $(\hat q_{1i}, \hat q_{2i})$ for each region. Assuming that
the proportions $M_i/M$ are approximately equal to the population
proportions for each region \citep{gelman} one can estimate the
difference in the fractions $q_1 - q_2$ by
\BEA
\label{frac_diff}
\sum_{i=1}^{16} \frac{M_i}{M} (\hat q_{1i} - \hat q_{2i}).
\EEA
Now we can apply MAXENT method for each region $i$ 
using frequency encoding $(z_{1i},z_{2i},z_{3i}) = (m_{1i}/M_i, m_{2i}/M_i, m_{3i}/M_i)$ 
and get corresponding estimates $(q^{[1]}(z_{1i}), q^{[1]}(z_{2i}), q^{[1]}(z_{3i}))$. 
The rightmost column in the table above shows the differences $q^{[1]}(z_{1i}) - q^{[1]}(z_{2i})$. 
Thus, the MAXENT estimate of the difference in the fractions $q_1 - q_2$ can be computed as in (\ref{frac_diff}) 
\BEA
q^{[1]}(z_1) - q^{[1]}(z_2) = \sum_{i=1}^{16} \frac{M_i}{M} (q^{[1]}(z_{1i}) - q^{[1]}(z_{2i})) = 0.149 > 0.
\EEA

\begin{table}
\centering
\caption{The regional distribution of the election data. Here Proportion approximates $M_i/M$, where $M_i$
is the sample length in each region, while $M=1447$ is the overall sample length; see \citep{gelman} for details. }
\begin{tabular}{lcccc|c}
\hline
Region &  Bush & Dukakis & Other & Proportion & $q^{[1]}(z_{1i}) - q^{[1]}(z_{2i})$
\\ \hline 
  Northeast, I &  0.298 &    0.617 &  0.085 &       0.032 & -0.404 \\
  Northeast, II &  0.500 &    0.478 &  0.022 &       0.032 & 0.070 \\
 Northeast, III &  0.467 &    0.413 &  0.120 &       0.115 & 0.093\\
  Northeast, IV &  0.464 &    0.522 &  0.014 &       0.048 & -0.180\\
     Midwest, I &  0.404 &    0.489 &  0.106 &       0.032 & -0.147\\
    Midwest, II &  0.447 &    0.447 &  0.106 &       0.065 & 0.0\\
   Midwest, III &  0.509 &    0.388 &  0.103 &       0.080 & 0.197\\
    Midwest, IV &  0.552 &    0.338 &  0.110 &       0.100 & 0.292\\
       South, I &  0.571 &    0.286 &  0.143 &       0.015 & 0.330\\
      South, II &  0.469 &    0.406 &  0.125 &       0.066 & 0.105\\
     South, III &  0.515 &    0.404 &  0.081 &       0.068 & 0.201\\
      South, IV &  0.555 &    0.352 &  0.093 &       0.126 & 0.296\\
        West, I &  0.500 &    0.471 &  0.029 &       0.023 & 0.084\\
       West, II &  0.532 &    0.351 &  0.117 &       0.053 & 0.255\\
      West, III &  0.540 &    0.371 &  0.089 &       0.086 & 0.266\\
       West, IV &  0.554 &    0.361 &  0.084 &       0.057 & 0.294\\
\end{tabular}
\label{tab10}
\end{table}

To see if the prediction of MAXENT is reliable (on average) here, the
same Bayesian decision model for these samples is set up, where first a
sample of $(q_1, q_2, q_3)$ is drawn from the Dirichlet distribution
with $\alpha_1 = \alpha_2 = \alpha_3 = 1$, and then using this sample as
category probabilities, categorical data sets of size $M$ are generated
with categories replaced by its frequency encodings. The process is
repeated and the average $\langle \overline{ K }_1\rangle$ from
(\ref{bat}) is computed via generating $10^3$ instances of
$\{q_k\}_{k=1}^3$ and $10^3$ categorical samples. For the present case
$\alpha_1 = \alpha_2 = \alpha_3 = 1$ and $n=3$, we have
$\langle\overline{K[q,q^{[0]}]}\rangle = 0.265$ from (\ref{bruno}). 

Now for $M>15$ we get that $\langle \overline{ K
}_1\rangle<\langle\overline{K[q,q^{[0]}]}\rangle$, i.e. the MAXENT
solution is reliable. For example, at $M=17$ we have $\langle
\overline{K}_{\rm Bayes}\rangle=0.046< \langle \overline{ K
}_1\rangle=0.116<\langle\overline{K[q,q^{[0]}]}\rangle=0.265$, where
$\langle \overline{K}_{\rm Bayes}\rangle$ refers to the Bayesian
(posterior mean) estimator (\ref{lid}). Already for $M=47$ predictions
of MAXENT are close to those of the optimal Bayesian estimator $\langle
\overline{K}_{\rm Bayes}\rangle=0.019< \langle \overline{K
}_1\rangle=0.034$. For the actual sample size $M=1447$, we get even closer results $\langle
\overline{K}_{\rm Bayes}\rangle=0.00087<\langle \overline{K }_1\rangle=0.014$.
Note from Table~\ref{tab10} that $M_i=M_{\rm South,\,I}=22$ is the minimal sample length in regions.
Hence all our MAXENT predictions for regions are reliable in the above sense.

To summarize the present real categorical data example, we saw that the
frequency encoding of the categorical variable allows to apply MAXENT.
The MAXENT estimator (\ref{first}) agrees with Bayesian estimator, and
is going to be reliable already for modest sample sizes $M>15$. For a
sufficiently large $M$ the average KL distance of the MAXENT estimator
gets close to that of the (optimal) Bayes estimator.

\section{Affine symmetry and the minimum entropy principle }
\label{minimum}

Above we focused on MAXENT estimators (\ref{s1}, \ref{first})
(the first empiric moment is fixed) or (\ref{s2}, \ref{first},
\ref{second}) (the first and second empiric moments are fixed). As
discussed around (\ref{nomo}), both (\ref{first}) and (\ref{second}) lead to
affine-invariant probabilities. We studied several
alternative constraints that do not have the full affine symmetry, i.e.
this symmetry is partial and relates to restriction on the parameters in
(\ref{nomo}). An example of this is constraining the
square-root (fractional) moment [cf.~(\ref{s1}, \ref{s2})]
\BEA
\label{s1/2}
&& q^{[1/2]}(z_k)=\frac{ e^{-\beta_{1/2} \sqrt{z_k} }}{\sum_{l=1}^n e^{-\beta_{1/2} \sqrt{z_l} } },\\
&& \sum_{k=1}^n q_k\sqrt{z_k}=\frac{1}{M}\sum_{u=1}^M \sqrt{Z_{u}},
\label{s1/22}
\EEA
where $\beta_{1/2}$ is determined from (\ref{s1/22}), and 
where we assumed $z_k>0$. For estimator (\ref{s1/2})  the symmetry
(\ref{nomo}) is kept under $g>0$ and $h=0$. We denote the corresponding
average KL distances by $\langle\overline{K^{}}_{1/2}\rangle$.

Let us now compare two different MAXENT estimators each one employing
its own constraint; e.g. we compare (\ref{s1}) with (\ref{s1/2}).  We
saw from extensive numeric simulations that whenever these constraints
have different degrees of the affine symmetry, then the estimator having
the largest symmetry provides a smaller average KL distance. A
particular example of this general relation is:
\BEA
\langle\overline{K^{}}_{1}\rangle<\langle\overline{K^{}}_{1/2}\rangle,
\label{oko} 
\EEA
which was verified on parameters of 
Tables \ref{tab0}--\ref{tab2}.

Recall that Refs.~\citep{good_entropy,christ,minent} proposed the minimum
entropy principle: when comparing two possible contraints to be employed
in the maximum entropy method, then it is preferable to use the one that
provides the smaller (maximized) entropy. The heuristic motivation of the
principle is that it avoids overfitting by not insisting too much on the
entropy maximization. This principle was motivated
via the minimum description length in \citep{mdl}. 

We ask whether in cases similar to (\ref{oko}) we can compare the
average entropies, i.e. for (\ref{oko}) we compare
$\langle\overline{S[q^{[1]}(z_k)]}\rangle$ and
$\langle\overline{S[q^{[1/2]}(z_k)]}\rangle$, where the averages are
defined as in (\ref{bat}). In all cases we were able to check, relations
similar to (\ref{oko}) are accompanied by the result that the constraint
which provide a smaller average KL distance (i.e. a better costraint) also has a smaller average
entropy, e.g. 
\BEA
\label{oko2} 
\langle\overline{S[q^{[1]}(z_k)]}\rangle <
\langle\overline{S[q^{[1/2]}(z_k)]}\rangle. 
\EEA
The theoretical origin of this relation between the average KL distance and the average (maximized)
entropy is not yet clear. 
Here is a concrete numerical example that illustrates (\ref{oko}, \ref{oko2}). For parameters of Table \ref{tab01}
we noted for $M=55$: 
$\langle\overline{K^{}}_{1}\rangle=1.756 <\langle\overline{K^{}}_{1/2}\rangle=1.758$ and
$\langle\overline{S[q^{[1]}(z_k)]}\rangle =4.006<
\langle\overline{S[q^{[1/2]}(z_k)]}\rangle=4.008$.
\end{document}